\documentclass{article}
\usepackage{amsthm}
\usepackage{mdframed}
\usepackage{amsmath}
\usepackage{amssymb}
\usepackage{authblk}
\usepackage[utf8]{inputenc}
\usepackage{graphicx}
\usepackage{enumerate}
\usepackage{blkarray}
\usepackage{color}
 \usepackage{comment}
\usepackage{subfigure}

\usepackage{a4wide}
\usepackage{graphics} 
  \usepackage{epstopdf}

  \usepackage{hyperref}
\hypersetup{
    unicode=false,          
    pdftoolbar=true,        
    pdfmenubar=true,        
    pdffitwindow=false,     
    pdfstartview={FitH},    
    pdftitle={My title},    
    pdfauthor={Author},     
    pdfsubject={Subject},   
    pdfcreator={Creator},   
    pdfproducer={Producer}, 
    pdfkeywords={keywords}, 
    pdfnewwindow=true,      
    colorlinks=true,       
    linkcolor=blue,          
    citecolor=blue,        
    filecolor=magenta,      
    urlcolor=cyan           
}
\def\x{\textbf{x}}
\def\f{\bar{f}}

\title{Evolutionary dynamics with random payoff matrices}
\author[1]{Manh Hong Duong\thanks{h.duong@bham.ac.uk. The corresponding author}}
\author[2]{The Anh Han\thanks{T.Han@tees.ac.uk}}
\affil[1]{School of Mathematics,
University of Birmingham, UK.}
\affil[2]{ School of Computing, Engineering and Digital Technologies, Teesside University, UK}
\date\today

\begin{document}

\maketitle
\begin{abstract}
Uncertainty, characterised by randomness and stochasticity, is ubiquitous in applications of evolutionary game theory across various fields, including biology, economics and social sciences. The uncertainty may arise from various sources such as fluctuating environments, behavioural errors or incomplete information. Incorporating uncertainty into evolutionary models  is essential for enhancing their relevance to real-world problems. In this perspective article, we survey the relevant literature on evolutionary dynamics with random payoff matrices, with an emphasis  on continuous models. We also pose challenging open problems for future research  in this important area.   
\end{abstract}
\section{Introduction}
In traditional evolutionary game theory, players---whether they are organisms, strategies, or agents--- interact according to defined rules, with payoffs reflecting the benefits or costs tied to different strategies \cite{sigmund:2010bo,hofbauer1998evolutionary,nowak:2006bo}. These payoffs are usually deterministic and derived from  interactions between strategies, where  more successful strategies tend to reproduce more quickly and ultimately dominate over time within a population.

However, in many real-world scenarios, environmental fluctuations and incomplete information render payoffs unpredictable ~\cite{fudenberg:1992bv, gross2009generalized,GokhalePNAS2010, Galla2013,DuongHanJMB2016,hilbe2018indirect,wang2024evolution}. For example, the payoff for a predator strategy may vary due to changes in prey abundance or the presence of competing predators. Therefore, it is of vital importance for the applicability of evolutionary game theory to integrate randomness and uncertainty into the evolutionary models. This need has given rise to the concept of evolutionary games with random payoffs.

In evolutionary games with random payoffs, payoffs are represented as random variables drawn from specific probability distributions instead of being fixed values. Each encounter between players yields a payoff that can differ each time, reflecting the randomness of the environment. For example, payoffs could follow a normal distribution, uniform distribution, or any other suitable distribution based on the model requirements. The dynamics of evolutionary games with random payoffs can differ significantly from those with fixed payoffs. The random nature of payoffs can lead to diverse population dynamics. Under different distributions, populations may converge to certain strategies in some cases or continually oscillate in others. Analyzing games with random payoffs introduces complexity as the traditional approaches may not apply straightforwardly, and demands the development of new methods with tools from various fields such as game theory, dynamical systems, probability theory, as well as computer simulations. One of the challenges is deriving meaningful insights from complex and sometimes counterintuitive outcomes.

Over the last two decades, random games have found applications across various fields:
\begin{itemize}
\item \textbf{Biology}: Evolutionary games with random payoffs can model situations where organisms face fluctuating environments. For instance, in predator-prey dynamics, the availability of prey might vary unpredictably, impacting predator strategies and survival.

\item \textbf{Economics}: In markets, payoffs often fluctuate due to factors like changing consumer preferences or economic conditions. Agents (firms, investors) might adopt mixed or robust strategies that hedge against these uncertainties, often using risk-averse approaches in their evolutionary game framework.

\item \textbf{Social Dynamics}: Random-payoff games can model social dilemmas, like cooperation in uncertain environments, where the benefit of cooperation may vary depending on factors outside of the players' control.
\end{itemize}
We refer the reader to~\cite{fudenberg:1992bv, gross2009generalized,GokhalePNAS2010, Galla2013,DuongHanJMB2016,hilbe2018indirect,wang2024evolution} again, as well as the recent survey articles \cite{traulsen2023future,leimar2023game} for further information.

In this article, we survey the relevant literature on evolutionary dynamics with random payoff matrices, focusing on continuous models. We group works on this topics into related and complementary perspectives. 

Evolutionary games with random payoffs have opened up rich avenues for understanding how randomness shapes strategic behaviour in natural, economic, and social systems, providing a more realistic framework for modeling complex adaptive systems. Nonetheless, many interesting and important questions  remain that we have yet to fully understand. In this regard, we also present several significant  open problems for future research. 

The rest of the paper is organized as follows.  Section \ref{sec: ESS} discusses evolutionary stable strategies in random games. Section \ref{sec: statistics} focuses on the statistics of the number of equilibria for  replicator-mutator dynamics. Section \ref{sec: fluctuating} reviews social dilemmas under fluctuating environments. Section \ref{sec: prediction} focuses on the dynamical properties of random games. In each section, we mathematically formulate open problems related to the respective  research direction for future work. The final section, Section \ref{sec: summary}, provides further discussions and a concise summary of open problems.
\section{Evolutionarily stable strategies in random games}
\label{sec: ESS}
Evolutionarily stable strategy (ESS), introduced by Maynard Smith and Price in 1973 \cite{SP73}, is a foundational concept in evolutionary game theory. This concept refines the notion of  Nash equilibrium (see below) and  has many applications in  fields such as ecology, economics and social sciences. 

Consider a population of $n$ different strategies. The interactions between the strategies are characterised by an $n\times n$ symmetric payoff matrix $A=(A_{ij})_{i,j=1,\ldots, n}$. A mixed strategy $p$ is a probability distribution on the set of pure strategies, that is $p=(p_1,\ldots, p_n)\in \Delta_n$ where $\Delta_n$ is the $n$-dimensional simplex
\[
\Delta_n=\{\x\in \mathbb{R}^n_+: \sum_{i=1}^n x_i=1\}.
\]
The payoff function, $A(p,q)$, between two mixed strategies, $p$ and $q$, is a bilinear form associated to the matrix $A$ and is given by
\[
A(p,q)=\sum_{i,j=1}^n A_{ij}p_iq_j. 
\]
A mixed strategy $p\in \Delta_n$ is an ESS if it satisfies the two conditions \cite{SP73}:
\begin{itemize}
    \item $A(p,p)\geq A(q,p)$ for all $q\in \Delta_n$,
    \item if $q\neq p$ satisfies $A(q,p)=A(p,p)$ then $A(q,q)<A(p,q)$.
\end{itemize}
The support of an ESS $p$ is defined as $\delta(p)=\{i: p_i>0\}$. Biologically, the existence of a mixed ESS and the size of its support provides useful insight into the co-existence and diversity of the strategies ($|\delta(p)|=1$ corresponds to monomorphism and $|\delta(p)|>1$ to polymorphism). Thus, when the payoff entries $A_{ij}$ are random variables, two natural questions arise:  what is the probability that an ESS exists, and how large is the size of the support of the ESS?
In a series of works~\cite{Haigh1988,haigh1989large,Haigh1990,hart2008evolutionarily}, the author addresses these questions for some specific and small values of $n$ and distributions of the payoff entries, while the paper \cite{kontogiannis2009support} investigates the case of large $n$ and broader classes of distributions. A remarkable result obtained in \cite{kontogiannis2009support} is that when payoff entries behave either as uniform, or normally distributed random variables, almost all ESSs have support sizes
of $o(n)$, where $n$ is the number of possible types for a player. With the aforementioned biological relevance of the concept of the support of an mixed ESS, the results of these work have implications for understanding the diversity of stable strategies within a population and gives insights into the relationship between the number of strategies and the robustness of an ESS. They also highlight how randomness influences population dynamics, particularly in terms of stability, coexistence, and the prevalence of certain strategies over others. More recently, intriguing connections between these papers to random quadratic optimization problems have been established \cite{pittel2018random,chen2021sparsity}.

\textbf{Open problem}: The assumption made in these papers that the payoff entries are  independent and identically distributed (iid) random variables is overly restrictive for real-world applications, where the individuals in a population are often interconnected. Can the results be generalised to accommodate cases where the payoff entries are non-iid random variables, particularly correlated ones? 

\section{Equilibria of random replicator-mutator dynamics}
\label{sec: statistics}
The replicator-mutator equation, which is a set of ordinary differential equations, describes the evolution dynamics of different strategies in a population, where selection and mutation are both captured. It is an well-established mathematical framework that integrates the unavoidable mutation observed in various  biological and social settings  \cite{traulsen:2009aa,rand2013evolution,zisis2015generosity,mcnamara2013towards,adami2016evolutionary,wang2020robust}. 
This framework  has been utilised   in many application domains, including   evolution of collective behaviours \cite{Imhof-etal2005,nowak:2006bo},  social networks dynamics \cite{Olfati2007}, language evolution
\cite{Nowaketal2001}, population genetics \cite{Hadeler1981}, and autocatalytic reaction networks
\cite{StadlerSchuster1992}. 

We consider an infinitely large population  consisting of $n$ different strategies $S_1,\cdots, S_n$. Their  frequencies are denoted, respectively,  by $x_1,\cdots, x_n$, where $\sum_{i=1}^n x_{i}=1$. These frequencies evolve overtime under the influence of selection and mutation. The interaction of the individuals in the population takes place in randomly selected groups of $d$ participants, that is, they play and obtain their fitness from $d$-player games. The fitness of a player is  calculated as average of the payoffs that they achieve from the interactions using a theoretic game approach. By means of mutation, individuals in the population  might change their strategy to another randomly selected strategy, given by the so-called mutation matrix: $Q=(q_{ji}), j,i\in\{1,\cdots,n\}$. Here, $q_{ji}$ stands for  the  probability of an $S_j$ individual changing its strategy to $S_i$, satisfying that
\begin{equation}
\label{eq: mutation matrix}
\sum_{j=1}^n q_{ji}=1, \quad \forall~ 1\leq i\leq n.    
\end{equation}
The replicator-mutator is given by \cite{Komarova2001JTB,Komarova2004, Komarova2010, Pais2012} 
\begin{equation}
\label{eq: RME}
\dot{x}_i=\sum_{j=1}^n x_j f_j(\x)q_{ji}- x_i \f(\x),\qquad  i=1,\ldots, n,
\end{equation}
where $\x = (x_1, x_2, \dots, x_n)$ denotes the vector of strategies, $f_i(\x)$ is the fitness of $i$-th strategy,  and $\f(\x)=\sum_{i=1}^n x_i f_i(\x)$ denotes the average fitness of the whole population.  
An equilibrium point of the replicator-mutation dynamics \eqref{eq: RME} is a point $\x\in [0,1]^n$ that satisfies
\begin{equation}
\label{eq: equilibria eqn2}
\sum_{j=1}^n x_j f_j(\x)q_{ji}- x_i \f(\x)=0, \quad i=1,\ldots, n.    
\end{equation}
In the replicator dynamics ($Q=I$), \eqref{eq: RME} reduces to
\[
\dot{x}_i=x_i(f_i(\x)- \f(\x)),\qquad  i=1,\ldots, n,
\]
while the equation that determines an internal equilibrium \eqref{eq: equilibria eqn2}, that is a point $\x$ in the interior of the unit cube $[0,1]^n$ (noting that in replicator dynamics, the vertices of this unit cube are obviously equilibria) is simplified to
\[
f_i(\x)- \f(\x)=0,\qquad  i=1,\ldots, n.
\]
The explicit formulas for the average payoffs $\{f_i\}_{i=1}^n$ depend on the type of the game that the players evolve. In symmetric games, a player’s payoff within a group does not depend on the ordering of its members, while in asymmetric games, it does. In both cases, the average payoffs are all multivariate polynomials in $x_1,\ldots, x_n$, thus finding an equilibrium of a $d$-player $n$-strategy evolutionary game using the replicator-mutation dynamics \eqref{eq: RME} is equivalent to finding a root in $[0,1]^n$ of the system of polynomial equations \eqref{eq: equilibria eqn2}. 
Accordingly, when the payoff entries are random variables, finding an equilibrium of the replicator-mutator dynamics \eqref{eq: RME} becomes finding a root in the $n$-dimensional unit cube of the system of random polynomials. This correspondence has been revealed and explored in a series of our works \cite{DuongHanDGA2016,DuongHanJMB2016,DuongTranHanDGA,DuongTranHanJMB,duong2019persistence,can2022expected,duong2023random}. Let $\mathcal{N}$ be the expected number of internal equilibria in $d$-player two-strategy games in the replicator dynamics. It has been known \cite{DuongHanDGA2016,DuongHanJMB2016,can2022expected}, that in symmetric games, when the payoff entries are iid Gaussian random variables then
\[
\mathbb{E}(\mathcal{N})=\sqrt{\frac{d-1}{2}}(1+o(1))~\text{as}~ d\rightarrow +\infty.
\]
Remarkably, by revealing that the random polynomial arising from asymmetric games is exactly the renowned the Kostlan-Shub-Smale random polynomial, we have recently shown that \cite{duong2023random}
\[
\mathbb{E}(\mathcal{N})=\frac{1}{2}\sqrt{d-1}\quad \forall~d
\]
if the payoff entries are iid Gaussian random variables. Furthermore, a universality phenomenon holds, which states that if the payoff entries are independent with mean $0$, variance $1$ and finite $(2+\varepsilon)$-moment for some $\varepsilon>0$. Then
\begin{equation*}
\mathbb{E}(\mathcal{N}_{d,2})=\frac{\sqrt{d-1}}{2}+O((d-1)^{1/2-c}),
\end{equation*}
for some $c>0$ depending only on $\varepsilon$. In \cite{DuongTranHanDGA, DuongHanDGA2020} we have partially extended these results to the replicator-mutator dynamics and the case of correlated payoff entries respectively. It is also worth mentioning related works ~\cite{huang:2010aa,huang2012impact} on fixation probabilities and the average population fitness of random games capturing random mutations.

\textbf{Open problems:} We have conjectured, supported by numerical evidence in \cite{duong2023random}, that universality results also apply to the class of random polynomials derived from symmetric evolutionary games. Can this conjecture rigorously be proven? 
\section{Social dilemmas under fluctuating environments}
\label{sec: fluctuating}
\begin{figure}
    \centering
    \includegraphics[scale=0.25]{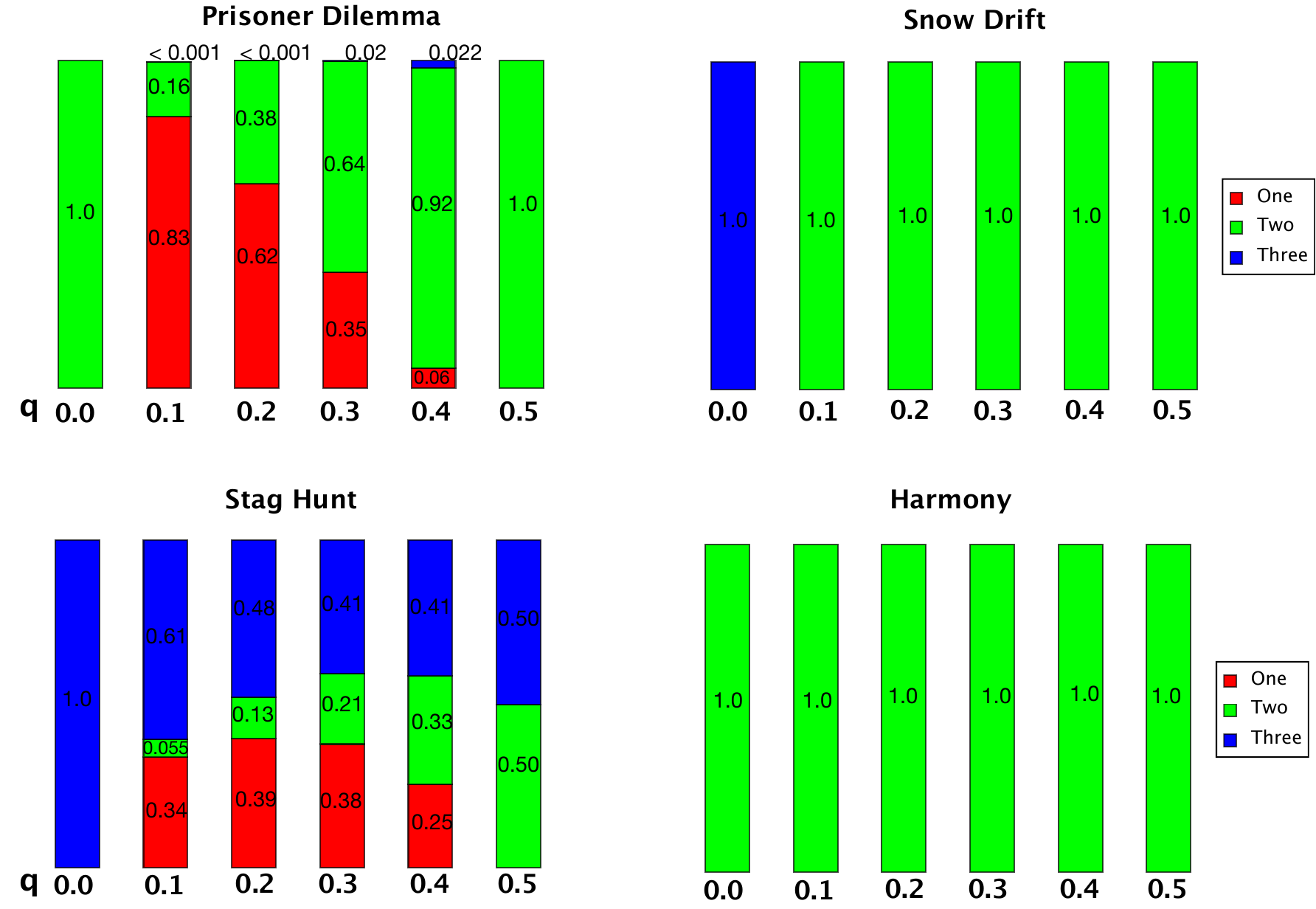}
    \caption{Probabilities of observing a certain number of equilibrium points for each social dilemma game, for different mutation strengths. Payoffs are drawn from uniform distributions. Reproduced from \cite{DuongHanDGA2020}.}
    \label{fig:social}
\end{figure}

Social dilemmas arise in many social, biological and economic contexts,  including multi-national conservation of natural resources, over-fishing and overgrazing of common property \cite{fatima2024learning}. In a simplified setting, social dilemmas are often modeled as a pairwise game with two strategies, cooperation (C) and defection (D),  with a general payoff matrix given by 
\[
\begin{blockarray}{ccc}
&C & D \\
\begin{block}{c(cc)}
  C&R & S \\
  D&T&P\\
  \end{block}
\end{blockarray}.
\]
Here $R$ (reward) and $P$ (punishment) are the payoffs for mutual cooperation and defection respectively, whereas $S$ (sucker) and $T$ (temptation) are the payoffs for cooperation by one player and defection by the other. The ordering among these payoff entries creates tension between individual and collective interests and leads to different outcomes. Important classes of social dilemmas include the Prisoner’s Dilemma game ($T > R > P > S$), the Snow-Drift  game ($T > R > S > P$), the Stag Hunt game ($R > T \geq P > S$) and the Harmonic Game ($R>T\geq P, R\geq S>P$).
The equilibrium equation of the replicator-mutator dynamics \eqref{eq: RME} for pairwise social dilemmas reads
\begin{equation}
\label{eq: SD}
(T+S-1)x^3+(1-T-2S+q(S-1-T))x^2+(S+q(T-S))x=0.    
\end{equation}
In many scenarios, the payoff entries are not fixed and fluctuate around certain known values, which are for example estimated through data analysis or given by domain experts. Hence studying the evolutionary outcomes of social dilemmas under the fluctuating environments has been of both practical and theoretical interest. In~\cite{DuongHan2021,chen2024number}, using the replicator-mutator dynamics \eqref{eq: SD}, we analyze the probability distribution of the number of equilibria as a function of the mutation strength for the four aforementioned social dilemmas, see Figure \ref{fig:social} for an illustration. Compared with the generic random games considered in Section \ref{sec: statistics}, in these works  the payoff entries fluctuate around the standard deterministic values that respect the orderings of the corresponding game. One can view this approach as a randomization of the concrete games to capture  uncertainty. By varying the type and amplitude of the noise, we reveal its influence on the evolutionary outcomes of the dynamics. In a similar setting, in \cite{AmaralRoyal2020,AmaralPRE2020} the authors show a rich spectrum of possible equilibria, highlighting the relationship
between heterogeneity and cooperation. From a more applied side, several  experimental studies  estimate the entries in the fitness matrix from in vitro data of interactions between microbial populations, for evolutionary game theory analysis \cite{turner1999prisoner,kaznatcheev2019fibroblasts,wolfl2022contribution}.  

\textbf{Open problems:} An interesting avenue for future research would be to analyze the randomization of established multi-player social dilemmas models (such as the multi-player Stag-Hunt and Snow-Drift Games, and the collective risk game) and to perform uncertainty quantification in evolutionary game theory.
\section{Dynamical properties of random games}
\label{sec: prediction}
The previous sections mainly focus on equilibrium properties of the replicator-mutator dynamics for multi-player two-strategy games. Now we consider two-player $n$-strategy games. Suppose $A=(A_{ij})_{i,j=1}^n$ is the payoff matrix. Then the replicator-mutator equation \eqref{eq: RME} becomes, for $i=1,\ldots, n$
\[
    \dot{x}_i=\sum_{j=1}^n x_j \Big(\sum_{i=1}^n A_{ji}x_i\Big)q_{ji}-x_i \sum_{i=1}^n\Big(\sum_{j=1}^n A_{ij}x_j\Big)x_i
\]
In the absence of mutation $Q=I$, we obtain the replicator dynamics:
\begin{align*}
\dot{x}_i&=x_i(f_i(\x)-\bar{f}(\x))\\&=x_i\Big(\sum_{j=1}^n A_{ij}x_j-\sum_{i=1}^n x_i\sum_{j=1}^n A_{ij}x_j\Big).   
\end{align*}
We are also interested in dynamical properties of this dynamics. As a dynamical system, it may exhibit different qualitative behaviours such as converging to a fixed point, having limit circles or becoming chaotic when $n\geq 3$ according to the Poincaré-Bendixon theorem. Together with equilibrium properties, the dynamical properties of the replicator-mutator dynamics often provide further useful insights into the underlying real-world systems. For instance, in the aforementioned applications of the replicator-mutator in language evolution and behaviour adoption in social networks \cite{Nowaketal2001,Komarova2001JTB, Komarova2004}, its limit cycles correspond to, respectively, oscillations in the dominance of the different grammars in the population, and oscillations of behavior preference, for example cycles in trends or fashions. Thus studying dynamical properties of the replicator-mutator dynamics is of both mathematical and practical interests. However, the replicator-mutator dynamics \eqref{eq: RME} is a nonlinear system with several parameters; hence, in general it is analytically intractable. 

In \cite{nowak:1993pv, mitchener2004chaos}, by sampling the random payoff entries and simulating the replicator-mutator dynamics, the authors demonstrate that the replicator dynamics for $n\geq 4$ can generate limit cycles and chaos for particular choices of fitness. Motivated by this prediction, in \cite{Pais2012}, the authors rigorously prove Hopf bifurcations for the replicator-mutator dynamics with $n\geq 3$ and characterize the existence of stable limit cycles for a class of  circulant fitness matrices. 

In recent years, the hybrid approach of combining sampling simulations and analytically proving the cyclic behaviour for the evolutionary dynamics has increasingly attracted attention from many researchers.   In \cite{park2020cyclic}, the authors consider the rock-paper-scissor games, which is an instance of the replicator-mutator dynamics described above with three strategies, with random payoff entries modeling the interactions occurring when a new mutant type arises. They show that the conditions needed for an interaction matrix to provide cyclic dominance are more restrictive than those for non-cyclic dominance, thus the chance to form cyclic dominance is small, explaining its rareness in nature. In a similar spirit and also for the rock-paper-scissor games, in \cite{kleshnina2021mistakes}, the authors show analytically that randomness in the payoff entries, interpreted as heterogeneity or individuals' behavioural mistakes, may break a cyclic relationship and lead to a stable equilibrium in pure or mixed strategies. In addition, in \cite{feng2022noise}, using tools in stochastic differential equations, the authors investigate how environmental noise may affect cyclic dynamical behavior of the rock–paper–scissors game. They deduce conditions for stochastic stability to determine whether environmental noise can induce a heteroclinic cycle. The aforementioned studies reveal the evolutionary significance of environmental stochastic fluctuations in biological populations with or without cyclic dominance. 

\textbf{Open problem:} The replicator-mutator dynamics for three or more strategies also arise naturally in recent works on multi-population models for AI governance and healthcare system \cite{alalawi2020toward, alalawi2024,bova2024both}. For instance, in the simplified model for AI governance \cite{alalawi2024}, the three populations consist of the regulators, AI users and the AI firms.  Similarly, for applications in healthcare, they are the patients, the private sectors and the public sectors \cite{alalawi2020toward}. Motivated by the aforementioned works on the rock-paper-scissors games, one interesting question is: can one derive dynamical properties such as limit cycles and bifurcations for these multi-population models?
\begin{mdframed}[backgroundcolor=green, linecolor=black,font={\sffamily},frametitle={\large{\underline{Summary of Open Problems}}}]
\begin{itemize}
    \item ESS for random games with non-iid payoff entries.
    \item Universality for the number of equilibria in symmetric random games.
    \item Randomization of multi-player social dilemmas.
    \item Uncertainty quantification in random games.
    \item Limit cycles for multi-population models.
\end{itemize}
\end{mdframed}
\section{Conclusion and Outlook}
\label{sec: summary}
In this paper, we have evidenced the role of integrating uncertainty into evolutionary game theory and discussed various important aspects of evolutionary dynamics with random payoff matrices. We  reviewed key works related to the foundational concept of evolutionarily stable strategies, social dilemmas under fluctuating environments as well as equilibrium and dynamical properties of random games. Additionally, we posed several challenging open problems for future research,  summarized in the summary table. We hope this perspective article will pave the way for further and deeper investigations into evolutionary dynamics with uncertainty in the years to come. 




\section*{Acknowledgment}
T.A.H. is supported by EPSRC (grant EP/Y00857X/1) and the Future of Life Institute. M.H.D is supported by EPSRC (grant EP/Y008561/1) and a Royal International Exchange Grant IES-R3-223047.
\bibliographystyle{abbrv}
\bibliography{ref}
\end{document}